\documentclass[aps,preprintnumbers,amsmath]{revtex4}
\usepackage{amssymb}
\usepackage{graphicx}
\usepackage{dcolumn}
\usepackage{bm}

\begin{document}
\title{Dependence of asymmetries for charge distribution with
respect \\ to the reaction plane on initial energy in heavy ion
collisions}

\author{V.A. Okorokov} \email{VAOkorokov@mephi.ru; okorokov@bnl.gov}
\affiliation{National Research Nuclear University "MEPhI",
Kashirskoe Shosse 31, 115409 Moscow, Russia}

\date{\today}

\begin{abstract}
In the paper two combinations of correlators are defined in order
to investigate the evolution of possible $\mathcal{P/CP}$
invariance violation in strong interactions with initial energy
for heavy ion collisions. These combinations correspond to
absolute and relative asymmetry of distribution of electrically
charge particles with respect to the reaction plane in heavy ion
collisions. Energy dependence of parameters under study was
derived from data of STAR and ALICE experiments. Significant
decreasing both absolute and relative asymmetry is observed at
energies $\sqrt{s_{NN}} < 20$ GeV. This feature agrees
qualitatively with other results of stage-I beam energy scan
program in STAR experiment. General behavior of dependence of
absolute asymmetry on initial energy agree reasonably with
behavior of similar dependence of Chern--Simons diffusion rate
calculated at different values of external Abelian magnetic field.
The observed behavior of parameters under study vs energy can be
considered as indication on possible transition to predominance of
hadronic states over quark-gluon degrees of freedom in the mixed
phase created in heavy ion collisions at intermediate energies.


\end{abstract}

\maketitle

\section{Introduction}\label{intro}
\hspace*{0.5cm}The vacuum of quantum chromodynamics (QCD) is a
very complicated matter, made of strongly interacting fermionic
and gauge fields, with rich geometry structure. This structure can
corresponds to the fractal geometry
\cite{Okorokov-arXiv-1206.0444-2012}. Hypothesis concerned on
complex and highly irregular geometry of QCD vacuum in Euclidean
space does not contradict both lattice calculations
\cite{Ilgenfritz-NPPS-B153-328-2006,Weinberg-XXIV-2006} and
fractal-like approximation of structures in distributions of
topological charge density \cite{Horvath-PR-D68-114505-2003}.
There is a fundamental interrelation between geometry and
fundamental properties of QCD Lagrangian. Vacuum of QCD as
non-Abelian gauge theory contains topologically non-trivial
configurations of gauge fields which deeply relate with
$\mathcal{P/CP}$ invariance of strong interactions. The
non-trivial topology of QCD vacuum opens the possibility for
existence of metastable domains which possess of various
properties with respect to the discrete $\mathcal{P/CP}$
symmetries. According to predictions of the theory at finite
temperature \cite{Lee-PRD-8-1226-1973} decays of such domains or
classical transitions (sphalerons) between them in deconfinement
phase of color charges with restored chiral symmetry can result in
local topologically induced violation of $\mathcal{P/CP}$
invariance in strong interactions -- $l\mathcal{TIP}$ effect. The
one of possible mechanisms for manifestation of such domains and
local $\mathcal{TIP}$ violation experimentally is the
azi\-mu\-thal asymmetry of distribution of electric charges in
final state driven by an external Abelian magnetic field -- the so
called chiral magnetic effect -- $\mathcal{CME}$
\cite{Kharzeev-NPA-803-227-2008}. Experimentally charge-separation
effect in relativistic heavy ion collisions was observed by STAR
collaboration at RHIC for the first time
\cite{Abelev-PRL-103-251601-2009} and was confirmed in further by
ALICE collaboration at LHC \cite{Abelev-arXiv-1207.0900-2012}. As
indicated above deconfinemet state of color charges and restored
chiral symmetry are essential for the experimental manifestation
of possible local strong $\mathcal{P/CP}$ violation, i.e. for
$\mathcal{CME}$. The former is needed to separate (anti-)quarks
with opposite electric charges by a distance larger than nucleon
size. The restored chiral symmetry is required because charge
separation is possible at conserved chirality only. Therefore the
investigation of energy dependence of $\mathcal{CME}$ via
definition of convenient parameters and analysis of its
corresponding dependence is important challenge for study of
strongly interacting matter under extreme conditions and search
for transition domain from prevalence of hadronic states to
domination of deconfinement phase of quarks and gluons. At present
the gauge-string duality seems one of the most powerful and
promising tools for study of various aspects of heavy ion
collisions in strong coupling regime \cite{Gubser-FP-43-140-2013}.
Thus the some approaches based on the gauge-string duality can be
useful for qualitative study of electric charge separation with
respect to the reaction plane which is non-perturbative
phenomenon.

The paper is organized as follows. In Sec.2, definitions of
collective variables (correlators) and asymmetries are described.
The Sec.3 devotes discussion of energy dependence of introduced
asymmetries, comparison of its experimental values with
approximations based on equation for Chern--Simons diffusion rate
at various strengths of external Abelian magnetic field. Some
final remarks are presented in Sec.4.

\section{Method and variables}\label{sec:2}
\hspace*{0.5cm}With taking into account possible local strong
$\mathcal{P/CP}$ violation the invariant distribution of final
state particles with certain sign of electric charge $\alpha$
($\alpha = +,-$) can be written as following
$$
E\frac{\textstyle d^{3}N_{\alpha}}{\textstyle
d\vec{p}}=\frac{\textstyle 1}{\textstyle 2\pi} \frac{\textstyle
d^{\,2}N_{\alpha}}{\textstyle
p_{\footnotesize{\perp}}dp_{\footnotesize{\perp}}dy}\biggl[1+\sum\limits_{n=1}^{\infty}2\left\{v_{n,\alpha}\cos
\left(n \Delta \phi\right)+a_{n,\alpha}\sin\left (n \Delta \phi
\right)\right\}\biggr].
$$
Here $\Delta \phi \equiv \phi-\Psi_{\mbox{\footnotesize{RP}}}$,
$\phi$ is an azimuthal angle of particle under study,
$\Psi_{\mbox{\footnotesize{RP}}}$ -- azimuthal angle of reaction
plane, $v_{n,\alpha}$ -- collective flow of $n$-th order, the
parameters $a_{n,\alpha}$ describe the effect of $\mathcal{P/CP}$
violation. The expectation value of $a_{n,\alpha}$ will vanish if
the one-particle distributions are averaged over event sample
because the distributions are averaged over many domains and
direction of charge separation changes event by event due to
random sign of the Chern--Simons number
($N_{\mbox{\footnotesize{CS}}}$) of the local domain
\cite{Kharzeev-NPA-803-227-2008,Abelev-PRL-103-251601-2009}.
Therefore it should be emphasized that the $\mathcal{CME}$ is the
collective effect and its investigation is possible via
correlation analysis only. According to the theory, the correlator
contained the contribution of possible $\mathcal{P/CP}$ violation
effect only is given by equation $\langle
\mathbf{K}_{n,\alpha\beta}^{\mbox{\scriptsize{T}}}\rangle =
\langle a_{n,\alpha}a_{n,\beta}\rangle$, where $\alpha, \beta$ --
the electric charge signs of secondary particles. Only the first
harmonic coefficient was analyzed so far because as expected the
$a_{1,\alpha}$ accounts for most of the effect under study
\cite{Abelev-PRL-103-251601-2009}. Thus the following notation
$\langle \mathbf{K}_{1,\alpha\beta}^{\mbox{\scriptsize{T}}}\rangle
\equiv \langle
\mathbf{K}_{\alpha\beta}^{\mbox{\scriptsize{T}}}\rangle$ is used
below. Experimental correlator had been proposed in
\cite{Voloshin-PRC-70-057901-2004} and it had been defined as
following $\langle
\mathbf{K}_{\alpha\beta}^{\mbox{\scriptsize{E}}}\rangle = \langle
\cos\left(\phi_{\alpha}+\phi_{\beta}-2\Psi_{\mbox{\footnotesize{RP}}}\right)\rangle$.
This observable is sensitive to the effect of possible local
strong $\mathcal{P/CP}$ violation and measures the charge
separation with respect to the reaction plane. Both theoretical
and experimental correlators are averaged over pairs of particles
under study in event and over all events in sample. It should be
mentioned that correlators defined above are $\mathcal{P}$-even
quantities therefore the $\langle
\mathbf{K}_{\alpha\beta}^{\mbox{\scriptsize{E}}}\rangle$ may
contains contributions from background effects unrelated to
possible local strong $\mathcal{P/CP}$ violation. Theoretical and
experimental correlators are related by following equation
\begin{equation}
\label{eq:1} \langle
\mathbf{K}_{\alpha\beta}^{\mbox{\scriptsize{E}}}\rangle =
\left[\langle
v_{1,\alpha}v_{1,\beta}\rangle_{\mbox{\scriptsize{even}}} + \Delta
\Phi \right] -\langle
\mathbf{K}_{\alpha\beta}^{\mbox{\scriptsize{T}}}\rangle.
\end{equation}
Here $v_{1,\alpha}$ is the directed flow parameter, $\Delta \Phi
\equiv
\Phi_{\mbox{\footnotesize{in}}}-\Phi_{\mbox{\footnotesize{out}}}$
-- the difference between background contribution of the in-plane
correlations $(\Phi_{\mbox{\footnotesize{in}}})$ and background
contribution of the out-of-plane correlations
$(\Phi_{\mbox{\footnotesize{out}}})$. Dependence of $v_{1}$ on
pseudorapidity can be generally separated into rapidity-odd and
rapidity-even component. The experimental contribution given by
odd part of the term $\langle v_{1,\alpha}v_{1,\beta}\rangle$ on
r.h.s. of (\ref{eq:1}) be neglected
\cite{Abelev-PRL-103-251601-2009,Voloshin-PRC-70-057901-2004,Okorokov-arXiv-0908.2522-2009}
because usual directed flow averages to zero at rapidities
symmetric with respect to midrapidity region which is considered
in most relativistic heavy ion experiments
\cite{Abelev-PRL-103-251601-2009,Abelev-arXiv-1207.0900-2012}. The
rapidity-even $v_{1}$ is originated from the dipole asymmetry of
the nuclear overlap which is result of fluctuations of the initial
geometry
\cite{Gardim-PRC-83-064901-2011,Teaney-PRC-83-064904-2011}. The
analysis of experimental data at $\sqrt{s_{NN}}=200$ GeV supports
the existence of a sizable rapidity-even $v_{1}$
\cite{Luzum-PRL-106-102301-2011}. These results were confirmed by
further investigations at the same energy
\cite{Pandit-arXiv-1211.7162} as well as for highest available
collision energy $\sqrt{s_{NN}}=2.76$ TeV
\cite{Aad-PRC-86-014907-2012,Retinskaya-PRL-108-252302-2012}. This
new type of directed flow results in nonzero contribution at
averaging on rapidity symmetric region.

In the framework of $\mathcal{CME}$ model
\cite{Kharzeev-NPA-803-227-2008} the theoretical correlator is
given by
\begin{eqnarray}
\langle \mathbf{K}_{\alpha\beta}^{\mbox{\scriptsize{T}}}\rangle &
=& \frac{\textstyle \pi^{2}}{\textstyle 16}\frac{\textstyle
\langle \Delta_{\alpha}\Delta_{\beta}\rangle}{\textstyle \langle
N_{\alpha}N_{\beta}\rangle} \\ \nonumber & =& \frac{\textstyle
\kappa \alpha_{\mbox{\scriptsize{S}}}}{\textstyle \langle
N_{\alpha}N_{\beta}\rangle} \biggl(\frac{\textstyle \pi
R}{\textstyle
2}\sum\limits_{f}q_{f}^{2}\biggr)^{2}\varepsilon_{\alpha\beta}\int\limits_{S_{\perp}}d^{2}x_{\perp}\xi_{\alpha\beta}(x_{\perp})
\int\limits_{\tau_{i}}^{\tau_{f}}d\tau \tau
\left[eB(\eta,\tau)\right]^{2}. \label{eq:2}
\end{eqnarray}
Here
$\varepsilon_{\pm\pm}=-0.5\varepsilon_{\pm\mp}=0.5,~\xi_{\pm\pm}(x_{\perp})
\equiv \sum_{i=+,-}\xi^{2}_{i}(x_{\perp}),~
\xi_{\pm\mp}(x_{\perp}) \equiv \prod_{i=+,-} \xi_{i}(x_{\perp}),$
$\alpha_{\mbox{\scriptsize{S}}}$ -- renormalized strong coupling
constant, $N_{\alpha}$ -- multiplicity of particles with electric
charge sign $\alpha$ in event, $R$ -- radius of colliding nuclei,
$q_{f}$ -- electric charge (in units of $e$) of quark with flavor
$f$, $\Delta_{\alpha}$ is the difference between total charges
with fixed sign $\alpha$ on each side of the reaction plane,
numerical coefficient $\kappa \sim 1$, $x_{\perp}$ -- position of
nucleon from one of the colliding nuclei in the plane
perpendicular to the beam axis, $S_{\perp}$ -- area of overlap
region of colliding nuclei in the transverse plane, $\tau$ --
proper time, $B$ -- strength of external Abelian magnetic field,
functions $\xi_{\pm}(x_{\perp})$ take into account the suppression
effect for transitions with change of Chern--Simons number in
various parts of the volume of created hot matter
\cite{Kharzeev-NPA-803-227-2008,Okorokov-arXiv-0908.2522-2009}.
Analytic approaches for (2) corresponded to various scenarios of
initial time and comparison with STAR results were discussed early
in elsewhere \cite{Okorokov-arXiv-0908.2522-2009}. The most
significant discrepancy between experimental and phenomenological
results observed for semi-central collisions
\cite{Okorokov-arXiv-0908.2522-2009} requests the additional
quantitative study and clear explanation. The STAR results for
azimuthal correlations of high $p_{\mbox{\footnotesize{\,T}}}$
particles with respect to the reaction plane
\cite{Adams-PRL-93-252301-2004} show that difference between such
correlations is largest for semi-central collisions. Possibly,
similar conclusion is valid for particles with moderate
$p_{\mbox{\footnotesize{\,T}}}$ too. Thus at qualitative level one
can suggests that second term $\Delta \Phi$ in r.h.s. of
(\ref{eq:1}) can influences on $\langle
\mathbf{K}_{\alpha\beta}^{\mbox{\scriptsize{E}}}\rangle$ and
results in the additional disagreement between experimental and
phenomenological data points.

In the framework of $\mathcal{CME}$ model
\cite{Kharzeev-NPA-803-227-2008} for ideal case of chiral limit
and an extremely large magnetic field the average square of charge
difference (in units of $e$) between opposite sides of the
reaction plane can be estimated as following
\begin{equation}
\langle Q^{2}\rangle = 4\langle
\Delta_{t}^{2}\rangle\biggl[\sum\limits_{f}|q_{f}|\biggr]^{2},
\label{eq:3}
\end{equation}
where $\Delta_{t}$ is the difference between the number of raising
and lowering of Chern--Simons number transition
\cite{Kharzeev-NPA-803-227-2008}. Let $\Delta
N_{\mbox{\footnotesize{CS}}}$ be a total change of Chern--Simons
number, which is dominated by the rate of corresponding change. In
this paper it has been suggested that at finite temperature ($T$)
the transition rate between vacuum states with different
$N_{\mbox{\footnotesize{CS}}}$ is dominated by transition rate due
to sphalerons only, without consideration of exponentially
suppressed transitions due to instantons. But it would be
mentioned that some additional study and justification may be
required for this approach, especially in temperature range
$T_{\mbox{\footnotesize{c}}} < T < 3T_{\mbox{\footnotesize{c}}}$
\cite{Ilgenfritz-PLB-325-263-1994}, where
$T_{\mbox{\footnotesize{c}}}$ -- temperature of phase transition
to the deconfinement state of color charges. It seems temperatures
reached experimentally in relativistic heavy ion collisions are in
the range indicated above for present and even future colliders.
Thus it is suggested that the rate of change of
$N_{\mbox{\footnotesize{CS}}}$ is the classical transition rate,
i.e. the Chern--Simons diffusion rate
($\Gamma_{\mbox{\footnotesize{CS}}}$). One can suggest $\Delta_{t}
\simeq \Delta N_{\mbox{\footnotesize{CS}}}$ because the most
probable transitions are those which connect two neighboring vacua
and transitions with simultaneously changing of absolute value of
Chern--Simons number larger than on unit are suppressed
\cite{Kharzeev-NPA-803-227-2008}. By definition the
$\Gamma_{\mbox{\footnotesize{CS}}}$ quality is the probability of
Chern--Simons number changing process to occur per unit
3-dimensional volume and per unit time: $
\Gamma_{\mbox{\footnotesize{CS}}} = \lim_{\mathcal{V} \to \infty}
{\mathcal{V}^{-1}}\langle [\Delta
N_{\mbox{\footnotesize{CS}}}]^{2}\rangle$, where $\mathcal{V}$ is
the 4-dimensional volume. The parameter
$\Gamma_{\mbox{\footnotesize{CS}}}$ has been intensively studied
in the framework of electroweak theory. The Chern--Simons
diffusion rate was computed for weak coupling domain. The
holographic AdS/CFT correspondence
\cite{Maldasena-ATMP-2-231-1998} makes it possible to derive this
parameter in $\mathcal{N}=4$ supersymmetric Yang--Mills (SYM)
plasma for limit of large number of colors ($N_{c} \to \infty$) in
the strong coupling regime \cite{Son-JHEP-0209-042-2002}. Then
with taking into account the sphaleron approach the Chern--Simons
diffusion rate in absent of external Abelian magnetic field can be
given by
$$
\Gamma_{\mbox{\footnotesize{CS}}}^{B=0}=f(\chi)T^{4},~~~
f(\chi)=\left\{
\begin{array}{lcl}
-k\chi^{5}\ln\chi, & \mbox{at} & \chi
\ll 1, \\
\chi^{2}/(256\pi^{3}), & \mbox{at} & \chi \gg 1.
\end{array}
\right.
$$
Here $\chi \equiv g^{2}N_{c}$ is the 't Hooft effective coupling,
$g$ -- gauge constant, the $\chi \ll 1$ corresponds to the weak
coupling domain and $\chi \gg 1$ -- to the strong coupling regime
\cite{Son-JHEP-0209-042-2002}. The latter corresponds to the
relativistic heavy ion collisions because the created matter seems
strongly correlated system. Numerical (large) coefficient $k$ is
of nonperturbative nature and can be found by a lattice
simulation. Estimations demonstrate that the $k \sim 10$ for SU(2)
\cite{Moore-PRD-58-045001-1998} and $k \sim 10^{2}$ for SU(3)
\cite{Moore-PLB-412-359-1997}, i.e. $k \sim 10^{N-1}$ for SU($N$)
with $N=2,3$. Thus $\Gamma_{\mbox{\footnotesize{CS}}}^{B=0}
\propto T^{4}$. In accordance with general definition of
$\Gamma_{\mbox{\footnotesize{CS}}}$ and (\ref{eq:3}) one can
obtain $\langle Q^{2}\rangle \propto
\Gamma_{\mbox{\footnotesize{CS}}}$. One the other hand in the most
ideal case under study, i.e. for contribution of the lowest Landau
level only, one can derive
$$
|\langle \Delta_{\pm}\Delta_{\pm}\rangle \mp  \langle
\Delta_{\pm}\Delta_{\mp}\rangle| = \varphi \langle [\Delta
N_{\mbox{\footnotesize{CS}}}]^{2}\rangle,~~\varphi \equiv
\frac{\textstyle 1}{\textstyle
2\mathcal{V}}\biggl[\sum\limits_{f}|q_{f}|\biggr]^{2}
\displaystyle \int\limits_{\mathcal{V}}
d^{4}x\left[\xi_{-}(x_{\perp}) \pm \xi_{+}(x_{\perp})\right]^{2}.
$$
The choice of difference of correlators in l.h.s. of the first
relation seems reasonable in order to corresponding quality be
positive, then
\begin{equation}
\langle \mathbf{K}_{\pm\pm}^{\mbox{\scriptsize{T}}}\rangle -
\langle \mathbf{K}_{\pm\mp}^{\mbox{\scriptsize{T}}}\rangle \propto
\langle [\Delta N_{\mbox{\footnotesize{CS}}}]^{2}\rangle \propto
\Gamma_{\mbox{\footnotesize{CS}}}. \label{eq:4}
\end{equation}

Taking into account the (\ref{eq:1}) one can define the following
quantity
\begin{equation}
A_{{\footnotesize{a}}}=-[\langle
\mathbf{K}_{\pm\pm}^{\mbox{\scriptsize{E}}}\rangle-\langle
\mathbf{K}_{\pm\mp}^{\mbox{\scriptsize{E}}}\rangle],\label{eq:5}
\end{equation}
which is absolute asymmetry for azimuthal distribution of electric
charges in final state and can be suggested for experimental
estimation of $\langle [\Delta
N_{\mbox{\footnotesize{CS}}}]^{2}\rangle$. The relative asymmetry
parameter for azimuthal distribution of electric charges with
respect to the reaction plane in event is defined as following
\begin{equation}
A_{{\footnotesize{r}}}=\frac{|\langle
\mathbf{K}_{\pm\pm}^{\mbox{\scriptsize{E}}}\rangle|-|\langle
\mathbf{K}_{\pm\mp}^{\mbox{\scriptsize{E}}}\rangle|}{\textstyle
|\langle
\mathbf{K}_{\pm\pm}^{\mbox{\scriptsize{E}}}\rangle|+|\langle
\mathbf{K}_{\pm\mp}^{\mbox{\scriptsize{E}}}\rangle|}.\label{eq:6}
\end{equation}

It should be noted that separation of possible $\mathcal{CME}$
signal from background effects is an important and difficult task
for $\langle
\mathbf{K}_{\alpha\beta}^{\mbox{\scriptsize{E}}}\rangle$ and
consequently for both $A_{{\footnotesize{a}}}$ and
$A_{{\footnotesize{r}}}$ quantities. The some physics backgrounds
for $\langle
\mathbf{K}_{\alpha\beta}^{\mbox{\scriptsize{E}}}\rangle$ were
discussed in experimental papers
\cite{Abelev-PRL-103-251601-2009,Abelev-arXiv-1207.0900-2012}.
Predictions of used event generators without incorporation of
$\mathcal{CME}$ effect contradict with dependence of $\langle
\mathbf{K}_{\alpha\beta}^{\mbox{\scriptsize{E}}}\rangle$ on
centrality for same-charge correlations
\cite{Abelev-PRL-103-251601-2009}. On the other hand these studies
do not take into account a number of important background effects.
First measurements \cite{Abelev-PRL-103-251601-2009} result in a
wide set of background studies (see, for example,
\cite{Bzdak-arXiv-0912.5050}). Some background correlations show
the the same sign for both same sign and opposite sign pairs of
particles
\cite{Abelev-PRL-103-251601-2009,Bzdak-PRC-83-014905-2011}. Thus
one can expect that the suggested definition (\ref{eq:5}) allows
to decrease contribution of some mutual backgrounds for absolute
asymmetry for azimuthal distribution of electrically charged
secondary particles. But the situation with total background
contribution in $\langle
\mathbf{K}_{\alpha\beta}^{\mbox{\scriptsize{E}}}\rangle$ is still
vague. Based on the experimental results
\cite{Abelev-PRL-103-251601-2009,Abelev-arXiv-1207.0900-2012,Mohanty-JPhysG-38-124023-2011,Wang-arXiv-1210.5498-2012}
and its interpretation for wide initial energy range as well as on
relations between absolute magnitudes of experimental signals and
some background correlations \cite{Abelev-PRL-103-251601-2009} one
can assume at qualitative level only that the sizable contribution
in $A_{{\footnotesize{a}}}$ and in $A_{{\footnotesize{r}}}$ will
be due to correlations driven by local $\mathcal{P/CP}$ violation
in strong interactions.

\section{Results: energy dependence for asymmetries}\label{sec:3}
\begin{figure*}
\includegraphics[width=16.0cm,height=16.0cm]{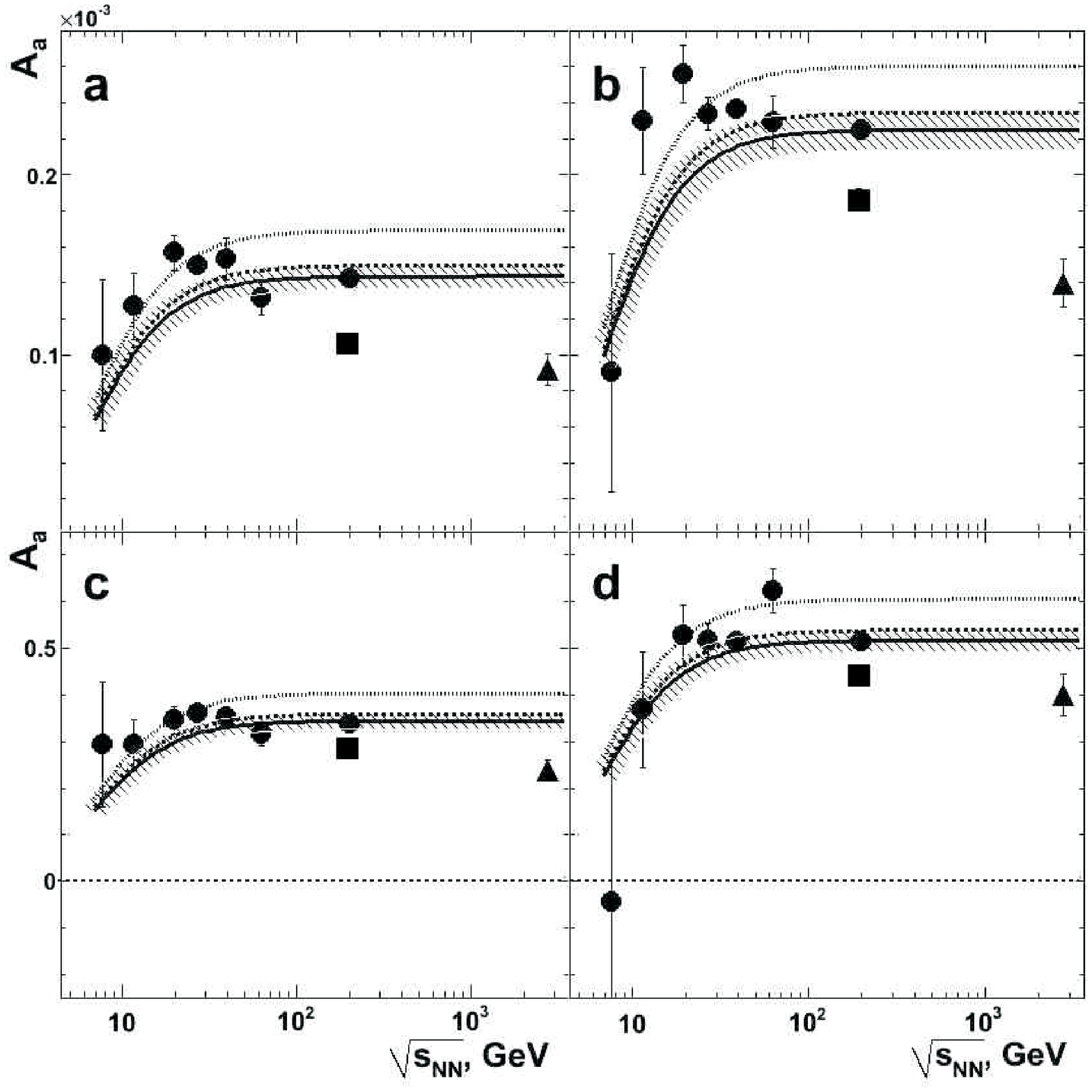}
\vspace*{8pt} \caption{Energy dependence of
$A_{{\footnotesize{a}}}$ for heavy ion collisions in various
centrality bins: a -- 20-30\%, b -- 30-40\%, c -- 40-50\% and d --
50-60\%. Experimental points are shown as following:
{\large$\bullet$} -- $\mbox{Au+Au}$, $\blacktriangle$ --
$\mbox{Pb+Pb}$ and $\blacksquare$ -- $\mbox{U+U}$. Absolute
asymmetry values are calculated based on experimental data from
\cite{Abelev-arXiv-1207.0900-2012,Mohanty-JPhysG-38-124023-2011,Wang-arXiv-1210.5498-2012}.
Smooth curves correspond to the energy dependence of Chern--Simons
diffusion rate in the strong coupling regime for external magnetic
field with strength $B=0$ (solid), $B=5T^{2}$ (dashed) and
$B=10T^{2}$ (dotted). The shaded areas for
$\Gamma_{\mbox{\footnotesize{CS}}}$ at $B=0$ are defined by
uncertainties of $T$ value at fixed initial energy due to errors
of parameters in analytic function described of
$T(\sqrt{s_{\footnotesize{NN}}})$ experimental dependence
\cite{Cleymans-JPhysG-32-S165-2006}.} \label{fig:1}
\end{figure*}

\hspace*{0.5cm}The choice of appropriate experimentally varied
parameters is crucial for study of features of phase diagram of
strongly interacting matter, in particular, for search for
critical point. It is important to have a control parameter whose
variation changes the chemical potential $\mu$ at which the
created matter crosses the transition region and freezes out. The
collision energy is an such parameter because varying the
collision energy has a large effect on $\mu$ at freezeout
\cite{Shuryak-book-2004}. Fig. \ref{fig:1} shows the energy
dependence of $A_{{\footnotesize{a}}}$ for semi-central events in
various bins of centrality for heavy ion beams. Absolute asymmetry
values are calculated based on experimental data from
\cite{Abelev-arXiv-1207.0900-2012,Mohanty-JPhysG-38-124023-2011,Wang-arXiv-1210.5498-2012}.
Collisions of moderate $\mbox{Cu}$ nuclei are characterized by
significantly larger values of $A_{{\footnotesize{a}}}$ than that
for heavy ions ($\mbox{Au}, \mbox{U}$) at corresponding energies.
It would be noted that for $\mbox{Au+Au}$ at
$\sqrt{s_{\footnotesize{NN}}}=62.4$ and 200 GeV absolute asymmetry
(\ref{eq:5}) was calculated based on the preliminary STAR
correlators for increased (full) statistics at former energy and
for Run 7 data sample at second initial energy
\cite{Wang-arXiv-1210.5498-2012}. These preliminary STAR results
allow to decrease the divergence of experimental points and get
slightly more smooth dependence $A_{{\footnotesize{a}}}$ on
$\sqrt{s_{\footnotesize{NN}}}$. The equation for Chern--Simons
diffusion rate for case of finite $B$ was derived in
\cite{Basar-arXiv-1202.2161-2012}. Background magnetic field
typically created in relativistic heavy ion collisions is
characterized by the strength $B \sim T^{2}$
\cite{Kharzeev-NPA-803-227-2008,Okorokov-arXiv-0908.2522-2009,Skokov-IJMPA-24-5925-2009}.
The regime of finite external magnetic field is investigated in
the paper and it is observed that the curves for $B=0$ and for
case $B \sim T^{2}$ coincide completely for all centrality bins
under consideration, i.e. for full centrality domain 20-60\% shown
at Fig. \ref{fig:1}. But relations (\ref{eq:4}) were obtained for
extremely strong magnetic field. Thus energy dependence for the
Chern--Simons diffusion rate are computed for external Abelian
magnetic field with various strengths as following
\cite{Basar-arXiv-1202.2161-2012}
$$
\Gamma_{\mbox{\footnotesize{CS}}}^{B \ne
0}=\Gamma_{\mbox{\footnotesize{CS}}}^{B=0}[1+\zeta^{2} /
(6\pi^{4})],~~~ \zeta \equiv B/T^{2}.
$$
It would be noted that at fixed $\zeta$ functional dependence of
$\Gamma_{\mbox{\footnotesize{CS}}}$ on $T$ is the same both for
$B=0$ and for general case of presence of finite external Abelian
magnetic field. Analytical function suggested in
\cite{Cleymans-JPhysG-32-S165-2006} for description of energy
dependence of chemical freeze-out temperature  agrees with
available experimental data quite reasonable for energies up to
$\sqrt{s_{\footnotesize{NN}}}=200$ Gev
\cite{Aggarwal-PRC-83-024901-2011} and for all centralities
\cite{Kumar-arXiv-1211.1350-2012} under study. Therefore the
analytic dependence $T(\sqrt{s_{\footnotesize{NN}}})$ from
\cite{Cleymans-JPhysG-32-S165-2006} is used for estimations of
$\Gamma_{\mbox{\footnotesize{CS}}}$ in energy domain under study.
The smooth curves at Fig. \ref{fig:1} are the energy dependence of
Chern--Simons diffusion rate in the strong coupling regime at
$B=0$ (solid), $B=5T^{2}$ (dashed) and $B=10T^{2}$ (dotted). The
norm for solid curves is the STAR point for $\mbox{Au+Au}$ at
$\sqrt{s_{\footnotesize{NN}}}=200$ GeV. The shaded bands for
Chern--Simons diffusion rate at $B=0$ are defined by uncertainties
of $T$ value at fixed initial energy due to errors of parameters
in analytic function described of
$T(\sqrt{s_{\footnotesize{NN}}})$ experimental dependence
\cite{Cleymans-JPhysG-32-S165-2006}.

As seen, phenomenological curves
$\Gamma_{\mbox{\footnotesize{CS}}}(\sqrt{s_{\footnotesize{NN}}})$
are in area caused by spread of $T$ values at fixed
$\sqrt{s_{\footnotesize{NN}}}$ for following wide range of
changing of strength of external Abelian magnetic field $B \leq
5T^{2}$ at any initial energies under study, moreover at
$\sqrt{s_{\footnotesize{NN}}} \lesssim 12$ GeV -- for all range of
changing of $B$ under considered on Fig. \ref{fig:1}. Functional
behavior of
$\Gamma_{\mbox{\footnotesize{CS}}}(\sqrt{s_{\footnotesize{NN}}})$
does not depend on $B$ in high energy domain at changing of $B$ on
order of magnitude at least. On the other hand the Chern--Simons
diffusion rate decreasing faster at $\sqrt{s_{\footnotesize{NN}}}
\leq 30$ GeV for larger strength of external Abelian magnetic
field. Possibly, the equation for Chern--Simons diffusion rate
without external $B$ is valid for any initial energies up to
$\sqrt{s_{\footnotesize{NN}}}=2.76$ TeV and for $\tau > 0.1$
fm/$c$ due to uncertainty area for
$\Gamma_{\mbox{\footnotesize{CS}}}(\sqrt{s_{\footnotesize{NN}}})$
at $B=0$ discussed above and $B$ values reached in heavy ion
collisions \cite{Okorokov-arXiv-0908.2522-2009}. Therefore further
decreasing of uncertainties of parameters which are used for
analytic approximation of $T(\sqrt{s_{\footnotesize{NN}}})$
behavior is essential for quantitative study of
$\Gamma_{\mbox{\footnotesize{CS}}}$ dependence on strength of
external Abelian magnetic field in nucleus-nucleus collisions.

One sees the some decreasing of $A_{{\footnotesize{a}}}$ for LHC
energy as compared to initial energy domain
$\sqrt{s_{\footnotesize{NN}}} \sim 100$ GeV, the absolute value of
decreasing growths for more peripheral events. Absolute asymmetry
of azimuthal distribution of electric charges in final state
$A_{{\footnotesize{a}}}$ goes down significantly with
$\sqrt{s_{\footnotesize{NN}}}$ decreasing for intermediate initial
energy domain 7.7 -- 20 GeV in semi-central heavy ion collisions
(Fig. \ref{fig:1}). For each centrality bin under study the
preliminary ex\-pe\-ri\-men\-tal results for
$A_{{\footnotesize{a}}}(\sqrt{s_{\footnotesize{NN}}})$ correspond
to the energy dependence of $\Gamma_{\mbox{\footnotesize{CS}}}$ at
$B=0$ on qualitative level with some enhancement of experimental
points over phenomenological curves in range
$\sqrt{s_{\footnotesize{NN}}} \sim 20-40$ GeV for events with
20-30\% (Fig. \ref{fig:1}a) and 30-40\% (Fig. \ref{eq:1}b)
centrality. Perhaps, for better description of
$A_{{\footnotesize{a}}}(\sqrt{s_{\footnotesize{NN}}})$ at
$\sqrt{s_{\footnotesize{NN}}} \sim 20-40$ GeV the influence of
external $B$ on $\Gamma_{\mbox{\footnotesize{CS}}}$ should be
taken into account. This suggestion agrees with dependence of $B$
on initial energy \cite{Okorokov-arXiv-0908.2522-2009} which
demonstrates that $B$ in intermediate energy range
$\sqrt{s_{\footnotesize{NN}}} \sim 20-40$ GeV reaches very large
values with respect to the strength of $B$ at
$\sqrt{s_{\footnotesize{NN}}} \sim 100$ GeV. The qualitative
agreement is observed between values of parameter (\ref{eq:5})
calculated via preliminary STAR results for $\mbox{Au+Au}$
collisions and phenomenological curves for any $B$ under
consideration. This feature can be considered as some evidence of
validity of equations (\ref{eq:3}) and (\ref{eq:4}) obtained at
ultimate strong external Abelian magnetic field for heavy ion
collisions. Decreasing of
$\Gamma_{\mbox{\footnotesize{CS}}}(\sqrt{s_{\footnotesize{NN}}})$
observed at $\sqrt{s_{\footnotesize{NN}}} < 20$ GeV should leads
to attenuation of $\mathcal{CME}$ and its manifestation on
experiment at intermediate energies. Taking into account
conditions which are essential for $\mathcal{CME}$ one can suppose
the following hypothesis. The changing of behavior of dependence
$A_{{\footnotesize{a}}}(\sqrt{s_{\footnotesize{NN}}})$ observed at
transition from high energy domain down to intermediate energy
range may be driven by predominance of hadronic colorless states
over quark-gluon deconfinement phase at
$\sqrt{s_{\footnotesize{NN}}} < 19.6$ GeV and, as consequence, by
decreasing of $\mathcal{CME}$. Thus behavior of experimental
parameter (\ref{eq:5}) vs collision energy agrees with qualitative
expectation for transition to the predominance of hadronic phase
in domain $\sqrt{s_{\footnotesize{NN}}} < 19.6$ GeV and with
decreasing of Chern--Simons diffusion rate at intermediate initial
energies.

\begin{figure*}
\includegraphics[width=16.0cm,height=16.0cm]{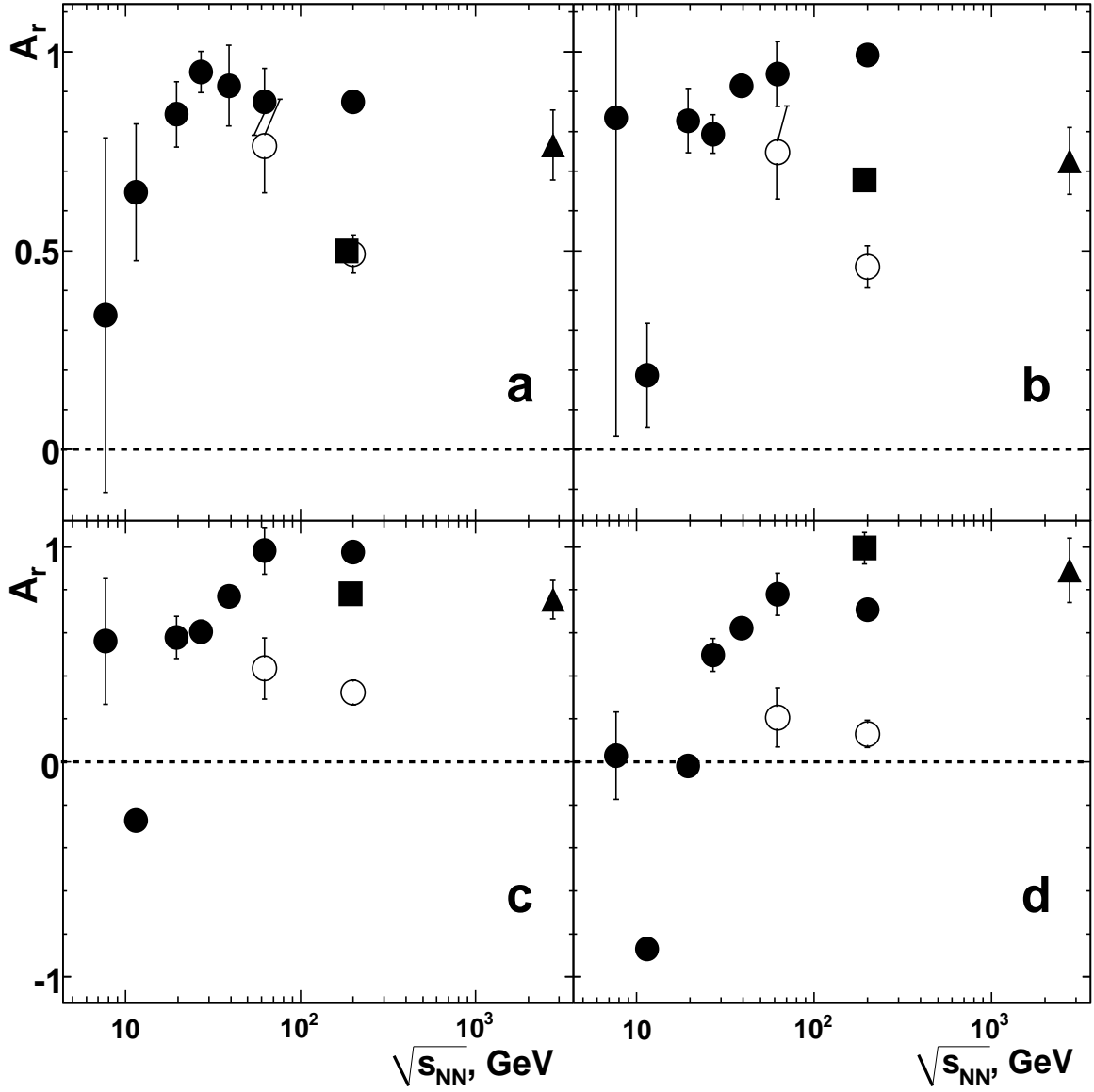}
\vspace*{8pt} \caption{Energy dependence of
$A_{{\footnotesize{r}}}$ for moderate and heavy nucleus-nucleus
collisions in various centrality bins: a -- 20-30\%, b -- 30-40\%,
c -- 40-50\% and d -- 50-60\%. Experimental points are shown as
following: {\large$\circ$} -- $\mbox{Cu+Cu}$, {\large$\bullet$} --
$\mbox{Au+Au}$, $\blacktriangle$ -- $\mbox{Pb+Pb}$ and
$\blacksquare$ -- $\mbox{U+U}$. Relative asymmetry values are
calculated based on experimental data from
\cite{Abelev-PRL-103-251601-2009,Abelev-arXiv-1207.0900-2012,
Mohanty-JPhysG-38-124023-2011,Wang-arXiv-1210.5498-2012}.}
\label{fig:2}
\end{figure*}

Fig. \ref{fig:2} demonstrates the energy dependence of
$A_{{\footnotesize{r}}}$ for semi-central (20-60\%) events for
various ion beam collisions. Relative asymmetry values are
calculated based on experimental data from
\cite{Abelev-PRL-103-251601-2009,Abelev-arXiv-1207.0900-2012,Mohanty-JPhysG-38-124023-2011,Wang-arXiv-1210.5498-2012}.
As well for Fig. \ref{fig:1} the dependence under study is some
improved by calculations of parameter (\ref{eq:6}) based on
increased STAR statistics at $\sqrt{s_{\footnotesize{NN}}}=62.4$
and 200 GeV. One sees the discrepancy of $A_{{\footnotesize{r}}}$
values for moderate and heavy nuclei at same energies increases
for more peripheral events. Behavior of
$A_{{\footnotesize{r}}}(\sqrt{s_{\footnotesize{NN}}})$ for heavy
ions changes dramatically at $\sqrt{s_{\footnotesize{NN}}} < 19.6$
GeV, especially for last three bins of centrality (Fig.
\ref{fig:2}b-d). The sharp decreasing of relative asymmetry
(\ref{eq:6}) for energies indicated above can be considered,
possibly, as evidence for (significant) amplification of hadronic
phase influence on experimental observables. The observed feature
of $A_{{\footnotesize{r}}}(\sqrt{s_{\footnotesize{NN}}})$ and
above suggestion agree with other STAR results obtained in the
framework of stage-I beam energy scan (BES-I) program at RHIC
\cite{Mohanty-JPhysG-38-124023-2011,Aggarwal-arXiv-1007.2613-2010}.

It should be emphasized that energy dependence of parameters
(\ref{eq:5}) and (\ref{eq:6}) obtained for any centralities from
most central (0-5\%) to most peripheral (70-80\%) events are
similar to them shown at Fig. \ref{fig:1} and Fig. \ref{fig:2}.

\section{Summary}\label{sec:4}
\hspace*{0.5cm}Two asymmetry parameters are introduced in order to
investigate the evolution of possible local strong
$\mathcal{P/CP}$ violation in heavy ion collisions with initial
energy. Dependence on $\sqrt{s_{\footnotesize{NN}}}$ has been
obtained for absolute ($A_{{\footnotesize{a}}}$) and relative
($A_{{\footnotesize{r}}}$) asymmetries based on the experimental
correlators. The the energy dependence of absolute asymmetry for
semi-central events in heavy ion collisions shows sharp decreasing
at $\sqrt{s_{\footnotesize{NN}}} < 19.6$ GeV, almost constant
behavior up to $\sqrt{s_{\footnotesize{NN}}} \simeq 200$ GeV with
some decreasing at further energy increasing. Dependence
$A_{{\footnotesize{a}}}(\sqrt{s_{\footnotesize{NN}}})$
qualitatively corresponds to the energy dependence of
Chern--Simons diffusion rate in $\mathcal{N}=4$ SYM plasma at
$N_{c} \to \infty$ in strong coupling regime. The main features of
$A_{{\footnotesize{r}}}$ energy dependence agree with behavior of
$A_{{\footnotesize{a}}}(\sqrt{s_{\footnotesize{NN}}})$ for wide
energy domain under study. Thus energy dependencies of suggested
asymmetry parameters with similar features at intermediate
$\sqrt{s_{\footnotesize{NN}}}$ indicate possibly on the beginning
of predominance of hadronic states versus phase of color degrees
of freedom in deconfinement state in the matter created in heavy
ion collisions with initial energies from domain
$\sqrt{s_{\footnotesize{NN}}} \lesssim 11.5 - 19.6$ GeV. Therefore
the asymmetry parameters and corresponding energy dependence seems
to be useful for study of $\mathcal{CME}$ and transition from
predominance of quark-gluon phase to hadronic one.

\section*{Acknowledgments}

I am grateful to G. Wang for discussions.

\newpage
\newpage

\end{document}